\documentclass{IEEETran}
\usepackage{epsfig}
\usepackage{booktabs} 
\usepackage{amsmath}
\usepackage{graphicx}
\usepackage{epstopdf}
\usepackage{graphics}
\usepackage{array}
\usepackage{float}
\usepackage{multirow}
\newtheorem{lem}{Lemma}
\newtheorem{theo}{Theorem}

\makeatletter

\providecommand{\tabularnewline}{\\}
\floatstyle{ruled}
\newfloat{algorithm}{tbp}{loa}
\providecommand{\algorithmname}{Algorithm}
\floatname{algorithm}{\protect\algorithmname}

\makeatother

\begin{document}

\title{Lightweight Hardware Architectures for Efficient Secure Hash Functions ECHO and Fugue}
\author{Mehran Mozaffari Kermani, Reza Azarderakhsh, Siavash Bayat-Sarmadi}
\IEEEspecialpapernotice{\thanks{Mehran Mozaffari Kermani is with Department of Computer Science and Engineering, University of South Florida, Tampa, FL 33620, Email: mehran2@usf.edu.} \thanks{Reza Azarderakhsh is with Department of ECE and Computer Science, Florida Atlantic University, Boca Raton, FL 14623, Email: razarderakhsh@fau.edu.}  \thanks{Siavash Bayat-Sarmadi is with Department of Computer Engineering, Sharif University of Technology, Tehran, Iran, Email: sbayat@sharif.edu.}}

\maketitle
\begin{abstract}
In cryptographic engineering, extensive attention has been devoted to ameliorating the performance and security of the algorithms within. Nonetheless, in the state-of-the-art, the approaches for increasing the reliability of the efficient hash functions ECHO and Fugue have not been presented to date. We propose efficient fault detection schemes by presenting closed formulations for the predicted signatures of different transformations in these algorithms. These signatures are derived to achieve low overhead for the specific transformations and can be tailored to include byte/word-wide predicted signatures. Through simulations, we show that the proposed fault detection schemes are highly-capable of detecting natural hardware failures and are capable of deteriorating the effectiveness of malicious fault attacks. The proposed reliable hardware architectures are implemented on the application-specific integrated circuit (ASIC) platform using a 65-nm standard technology to benchmark their hardware and timing characteristics. The results of our simulations and implementations show very high error coverage with acceptable overhead for the proposed schemes.
\end{abstract}

\section{Introduction}
Cryptographic hash functions take arbitrary-length inputs and generate fixed-length outputs. The output of hash function is then utilized to provide authentication and integrity for the transferred data. In this paper, due to the efficiency of the algorithms ECHO \cite{3} and
Fugue \cite{4} (which has been improved to Fugue 2.0), and the fact that these are inspired by the widely-utilized Advanced Encryption Standard (AES), we present their respective fault detection schemes. These AES-inspired hash functions (which have been part of the NIST competition) have received much attention in the literature. For instance, in \cite{66} and \cite{666}, differential and side-channel analysis attacks for ECHO are presented. Moreover, much effort has been put into developing high-performance and efficient hardware implementations of these algorithms, see, for instance, \cite{7}, \cite{9}, and \cite{10}. As discussed in \cite{11}, one important feature of these hash functions is that one can share some resources between the AES and these hash algorithms. Thus, low-complexity implementations are achieved.

Fault attacks pose serious threats to the implementations of the crypto-algorithms. Therefore, many fault detection schemes have been proposed to date for cryptographic and arithmetic entities, see, for instance, \cite{14}, \cite{15}, \cite{16}, \cite{18}, \cite{19}, \cite{20}, \cite{21}, \cite{22}, \cite{22q}, and \cite{22qq} for some examples. Nonetheless, to the best of our knowledge, the schemes for increasing the reliability of these algorithms have not been presented in the open literature. Effective fault detection schemes with minimal overhead on these algorithms are essential for achieving reliable hardware architectures.

The summary of our contributions is presented in the following.
\begin{itemize}
\item We have obtained new formulations for the predicted signatures of different transformations for hash algorithms, i.e., ECHO \cite{3} and
Fugue \cite{4}. The presented closed formulations are used for proposing high-performance and effective fault detection schemes.
\item Our simulation results show high fault detection capability for the proposed schemes for all the algorithms. This makes the proposed architectures reliable in practice.
\item We have used ASIC implementations to benchmark the hardware and timing characteristics of the proposed schemes. The high efficiency of the proposed schemes makes the proposed architectures suitable for high-performance applications.
\end{itemize}

\section{Preliminaries}
ECHO (presented by Benadjila \textit{et al.}) \cite{3} supports any hash output of length from 128 to 512 bits. The hash function ECHO takes a message and a salt as input. Although the output can be of any length from 128 to 512 bits, the four outputs for NIST competition were 224, 256, 384, and 512
bits. The ECHO algorithm with the output size ($H_{size}$) less than 256, i.e., $128\leq H_{size}\leq 256$, uses the compression function called Compress$_{512}$. However, for $257\leq H_{size}\leq 512$, the compression function is called Compress$_{1024}$ which is very similar to Compress$_{512}$ \cite{3}. More details are presented throughout the paper as needed.

In what follows, we explain the hash function Fugue (presented by the IBM) \cite{4}. Fugue-256 generates a 256-bit output $H$ for the message $M$ which is split into 32-bit blocks $m_i$, $1 \leq i \leq t$. The chaining value of Fugue-256 (denoted by $h$) is also split to 32-bit blocks denoted by $S_i$, $0 \leq i\leq 29$. The following transformation sequence is used for updating $h$ from $m_i$: TIX, ROR3, CMIX, SMIX, ROR3, CMIX, and SMIX (called one round $R$). The sequence ROR3, CMIX, SMIX is called a sub-round. Therefore, a round $R$ consists of the TIX transformation followed by two sub-rounds \cite{4}. More details are presented throughout the paper as needed.
\section{The Proposed Fault Diagnosis Approaches}
In what follows, for each of the algorithms presented in this paper, we propose respective fault detection schemes.

\subsection{ECHO}
An overview of the ECHO algorithm for $128\leq H_{size}\leq 256$ including the Compress$_{512}$ functions is presented in Fig. 1. As seen in Fig. 1, each of the $t$ Compress$_{512}$ functions gets the 128-bit salt, a $4\times4$ state of 128-bit entries, and the counter $C_i$, $1\leq i\leq t$ (used
to count the number of message bits being hashed). The first column of the state consists of four 128-bit values which construct the chaining variable of the previous Compress$_{512}$, i.e., $V_{i-1}=(v_{i-1}^0,v_{i-1}^1,v_{i-1}^2,v_{i-1}^3)$, $1\leq i\leq t$. The other three columns include the 128-bit blocks of the input message. Therefore, in total, there are $12\times t$ 128-bit message blocks to be processed to give the output (see Fig. 1).

As in seen Fig. 1, each Compress$_{512}$ consists of four different transformations, i.e., BIG.SubWords, BIG.ShiftRows, BIG.MixColumns, and BIG.Final. Each BIG.SubWords contains two AES rounds. The first transformation SubBytes which includes 16 S-boxes is the only nonlinear AES transformation. In the AES S-box, the irreducible polynomial of $M(x)=x^8+x^4+x^3+x+1$ is used to
construct the binary field $GF(2^8)$. Let $X\in GF(2^8)$ and $Y\in GF(2^8)$ be
the 8-bit input and output of each S-box, respectively. Then, the
S-box consists of a multiplicative inversion, i.e., $X^{-1}\in
GF(2^8)$, followed by an affine transformation to obtain $Y\in GF(2^8)$. Look-up tables (LUTs) and composite fields (polynomial basis, normal basis, mixed basis, and redundant-basis are among the approaches for this low-area implementation variant \cite{Hodjat,30-1,31-1,32-1}) are used to implement the S-boxes. In general, with composite field realizations, a transformation matrix first transforms a field element in the binary
field $GF(2^8)$ to the corresponding representation in the composite
fields $GF(2^8)$/$GF(((2^2)^2)^2)$. Then, a multiplicative inversion consisting of composite field operations in the sub-field $GF((2^2)^2)$ is performed. Finally, through an inverse transformation matrix, the inverted output is obtained. There have been a number of great research works for error detection of the S-boxes and for the sake of brevity, we do not discuss them.\begin{figure}[t]
\begin{center}
\epsfig{file=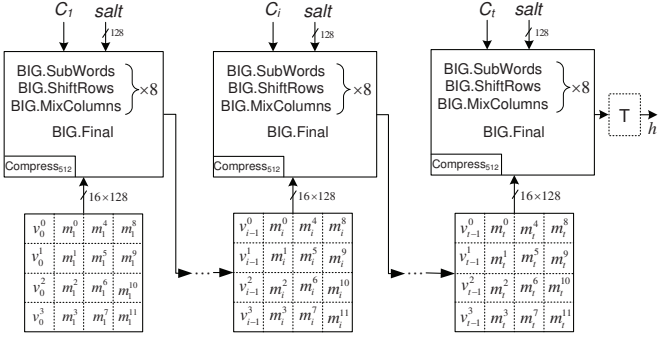,width=1\linewidth,clip=} \caption{The ECHO algorithm for $128\leq H_{size}\leq 256$ \cite{3}.}
\end{center}
  \end{figure}

The next transformation used in BIG.SubWords of ECHO is ShiftRows whose fault detection is straightforward and by re-wiring. Moreover, for the two final linear transformations, i.e., MixColumns and AddRoundKey, the 32-bit error indication flag $E_c=\sum_{r=0}^3({in}_{r,c}+k_{r,c}+{out}_{r,c})$, $0\leq c \leq 3$, can be used. It is noted that ${in}_{r,c}$, $k_{r,c}$, and ${out}_{r,c}$ are the input to MixColumns, the round key, and the output of AddRoundKey, respectively. This error indication flag can be compressed so that an $n$-bit, $1\leq n \leq 32$, error indication flag for these two transformations are achieved. Finally, after two rounds of the AES, the output of BIG.SubWords is derived.

Fault detection for the next transformation in ECHO, BIG.ShiftRows, is by permutation. As explained in the aforementioned explanation, the last transformation in BIG.Round, i.e., BIG.MixColumns, is an expansion of MixColumns of the AES. Specifically, the output state of BIG.SubWords (input state of BIG.MixColumns) is arranged as a 4-row, 64-column matrix. Then, each $4\times4$ sub-matrix is multiplied by the fixed MixColumns matrix. Therefore, we obtain the error indication flags of the BIG.MixColumns (B.MC) transformation for $j$ sub-matrices, $0\leq j \leq 15$, as follows
\begin{equation}
E{^j_c} (B.MC)=\sum_{r=0}^3({in}_{r,c}+{out}_{r,c}),\; \quad 4j\leq c \leq 4j+3,
\end{equation}where in the sub-matrices, ${in}_{r,c}$ and ${out}_{r,c}$ are the input and output of BIG.MixColumns, respectively, for which $0\leq r \leq 3$ and $0\leq c \leq 63$.

Finally, the BIG.Final transformation is performed as the last transformation in each Compress$_{512}$ (see Fig. 1) of ECHO. This transformation includes modulo-2 addition of the input state of the Compress$_{512}$ and the output state of the eighth BIG.MixColumns. We present the following lemma for obtaining the predicted parities of this transformation.
\begin{lem}
Let $M^j_i$, $0\leq j \leq 11$, be the 128-bit message blocks and $A^j_i$, $0\leq j \leq 15$, be the 128-bit outputs of the eighth BIG.MixColumns of the i$^{th}$ Compress$_{512}$ in Fig. 1. In addition, let $v^{j}_{i-1}$, $0\leq j \leq 3$, be the previous chaining values. Then, the predicted parities of $v^{j}_{i}$, $0\leq j \leq 3$ (the current chaining values), after performing the BIG.Final transformation is obtained as
\begin{equation}
\hat{P}(v^{j}_{i})=\sum_{j=0}^3 P(v^{j}_{i-1}+A^{4j}_i)+\sum_{j=0}^2 P(M^{4j}_i).
\end{equation}\end{lem}

\textit{Proof.}
\textit{According to \cite{3}, we have $v^{j}_{i}=\sum_{j=0}^3 v^{j}_{i-1}+ \sum_{j=0}^11 A^{j}_i+\sum_{j=0}^15 M^{j}_i$. Therefore, for the predicted parity we reach $\hat{P}(v^{j}_{i})=\sum_{j=0}^3 P(v^{j}_{i-1})+ \sum_{j=0}^11 P(A^{j}_i)+\sum_{j=0}^15 P(M^{j}_i)$ and after rearranging, the proof is complete.}

It is interesting to note that one can also obtain multiple parities for $v^{j}_{i}$ by applying the parity derivation function ($P$) to selected bits of the arguments $v^{j}_{i-1}+A^{4j}_i$ and $M^{4j}_i$.

\subsection{Fugue}
To propose a fault detection scheme for Fugue, we observe that the Fugue transformations can be divided into three types. The first type is the rotation transformations, i.e., ROR3, ROR14, and ROR15. The second category contains the two linear transformations TIX and CMIX. Finally, the last one is the nonlinear transformation SMIX.

Each Fugue round has the following sequence: TIX, ROR3, CMIX, SMIX, ROR3, CMIX, and SMIX. First, we propose the following theorem for the first three transformations TIX, ROR3, and CMIX in the round sequence. Then, we propose the fault detection scheme for the nonlinear transformation SMIX.
\begin{theo}
Let $\sigma_{S_i}=\sum_{i=0}^{29}S_i$ be the 32-bit result of modulo-2 additions of $S_i$, $0 \leq i\leq 29$ (called word-wide signature). Then, the predicted word-wide signature of the transformations sequence TIX, ROR3, and CMIX ($\hat{\sigma}_{TRC}$) in the Fugue round is obtained as
\begin{align}
\hat{\sigma}_{TRC}=\sigma_{S_i}+S_{24}.
\end{align}
\end{theo}
\textit{Proof.}
\textit{
For TIX, the following substitutions are performed: $S_{10}\leftarrow S_{10} + S_{0}$, $S_{0}\leftarrow m_{i}$, $S_{8}\leftarrow S_{8} + m_{i}$, and $S_{1} \leftarrow S_{1} + S_{24}$. Therefore, we have $\hat{\sigma}_{TIX}=\sigma_{S_i}+S_{10}+S_{10}+S_{0}+S_{0}+m_{i}+S_{8}+S_{8}+m_{i}+S_{1}+S_{1}+ S_{24}=\sigma_{S_i}+S_{24}$. The ROR3 transformation, which is just rotations three positions to right, does not change $\hat{\sigma}_{TIX}=\sigma_{S_i}+S_{24}$. Moreover, for CMIX, we have $S_{0}\leftarrow S_{0} + S_{4}$, $S_{1}\leftarrow S_{1} + S_{5}$, $S_{2}\leftarrow S_{2} + S_{6}$, $S_{15}\leftarrow S_{15} + S_{4}$, $S_{16}\leftarrow S_{16} + S_{5}$, and $S_{17}\leftarrow S_{17} + S_{6}$. Consequently, we reach $\hat{\sigma}_{CMIX}=\sigma_{S_i}+S_{0}+S_{0}+S_{4}+S_{1}+S_{1}+S_{5}+S_{2}+S_{2}+S_{6}+S_{15}+S_{15}+S_{4}+S_{16}+S_{16}+S_{5}+S_{17}+S_{17}+S_{6}=\sigma_{S_i}$. Therefore, one reaches $\hat{\sigma}_{TRC}=\sigma_{S_i}+S_{24}$ and the proof is complete.}

The nonlinear transformation SMIX in Fugue consists of two functions. The second one is the linear Super-Mix function. The Super-Mix function consists of multiplication of $S_0$-$S_3$ (as a 16-byte input vector) with the following $16\times16$ matrix $\boldsymbol{N}$ with hexadecimal entries to derive a 16-byte output
\begin{eqnarray}{\scriptsize{\boldsymbol N}=\left(\begin{array}{cccc}
1\hspace{5pt}4\hspace{5pt}7\hspace{5pt}1&\hspace{5pt}1\hspace{5pt}0\hspace{5pt}0\hspace{5pt}0&\hspace{5pt}1\hspace{5pt}0\hspace{5pt} 0\hspace{5pt}0&\hspace{5pt}1\hspace{5pt}0\hspace{5pt}0\hspace{5pt}0\\
0\hspace{5pt}1\hspace{5pt}0\hspace{5pt}0&\hspace{5pt}1\hspace{5pt}1\hspace{5pt}4\hspace{5pt}7&\hspace{5pt}0\hspace{5pt}1\hspace{5pt} 0\hspace{5pt}0&\hspace{5pt}0\hspace{5pt}1\hspace{5pt}0\hspace{5pt}0\\
0\hspace{5pt}0\hspace{5pt}1\hspace{5pt}0&\hspace{5pt}0\hspace{5pt}0\hspace{5pt}1\hspace{5pt}0&\hspace{5pt}7\hspace{5pt}1\hspace{5pt} 1\hspace{5pt}4&\hspace{5pt}0\hspace{5pt}0\hspace{5pt}1\hspace{5pt}0\\\vspace{5pt}
0\hspace{5pt}0\hspace{5pt}0\hspace{5pt}1&\hspace{5pt}0\hspace{5pt}0\hspace{5pt}0\hspace{5pt}1&\hspace{5pt}0\hspace{5pt}0\hspace{5pt} 0\hspace{5pt}1&\hspace{5pt}4\hspace{5pt}7\hspace{5pt}1\hspace{5pt}1\\
0\hspace{5pt}0\hspace{5pt}0\hspace{5pt}0&\hspace{5pt}0\hspace{5pt}4\hspace{5pt}7\hspace{5pt}1&\hspace{5pt}1\hspace{5pt}0\hspace{5pt} 0\hspace{5pt}0&\hspace{5pt}1\hspace{5pt}0\hspace{5pt}0\hspace{5pt}0\\
0\hspace{5pt}1\hspace{5pt}0\hspace{5pt}0&\hspace{5pt}0\hspace{5pt}0\hspace{5pt}0\hspace{5pt}0&\hspace{5pt}1\hspace{5pt}0\hspace{5pt} 4\hspace{5pt}7&\hspace{5pt}0\hspace{5pt}1\hspace{5pt}0\hspace{5pt}0\\
0\hspace{5pt}0\hspace{5pt}1\hspace{5pt}0&\hspace{5pt}0\hspace{5pt}0\hspace{5pt}1\hspace{5pt}0&\hspace{5pt}0\hspace{5pt}0\hspace{5pt} 0\hspace{5pt}0&\hspace{5pt}7\hspace{5pt}1\hspace{5pt}0\hspace{5pt}4\\\vspace{5pt}
4\hspace{5pt}7\hspace{5pt}1\hspace{5pt}0&\hspace{5pt}0\hspace{5pt}0\hspace{5pt}0\hspace{5pt}1&\hspace{5pt}0\hspace{5pt}0\hspace{5pt} 0\hspace{5pt}1&\hspace{5pt}0\hspace{5pt}0\hspace{5pt}0\hspace{5pt}0\\
0\hspace{5pt}0\hspace{5pt}0\hspace{5pt}0&\hspace{5pt}7\hspace{5pt}0\hspace{5pt}0\hspace{5pt}0&\hspace{5pt}6\hspace{5pt}4\hspace{5pt} 7\hspace{5pt}1&\hspace{5pt}7\hspace{5pt}0\hspace{5pt}0\hspace{5pt}0\\
0\hspace{5pt}7\hspace{5pt}0\hspace{5pt}0&\hspace{5pt}0\hspace{5pt}0\hspace{5pt}0\hspace{5pt}0&\hspace{5pt}0\hspace{5pt}7\hspace{5pt} 0\hspace{5pt}0&\hspace{5pt}1\hspace{5pt}6\hspace{5pt}4\hspace{5pt}7\\
7\hspace{5pt}1\hspace{5pt}6\hspace{5pt}4&\hspace{5pt}0\hspace{5pt}0\hspace{5pt}7\hspace{5pt}0&\hspace{5pt}0\hspace{5pt}0\hspace{5pt} 0\hspace{5pt}0&\hspace{5pt}0\hspace{5pt}0\hspace{5pt}7\hspace{5pt}0\\\vspace{5pt}
0\hspace{5pt}0\hspace{5pt}0\hspace{5pt}7&\hspace{5pt}4\hspace{5pt}7\hspace{5pt}1\hspace{5pt}6&\hspace{5pt}0\hspace{5pt}0\hspace{5pt} 0\hspace{5pt}7&\hspace{5pt}0\hspace{5pt}0\hspace{5pt}0\hspace{5pt}0\\
0\hspace{5pt}0\hspace{5pt}0\hspace{5pt}0&\hspace{5pt}4\hspace{5pt}0\hspace{5pt}0\hspace{5pt}0&\hspace{5pt}4\hspace{5pt}0\hspace{5pt} 0\hspace{5pt}0&\hspace{5pt}5\hspace{5pt}4\hspace{5pt}7\hspace{5pt}1\\
1\hspace{5pt}5\hspace{5pt}4\hspace{5pt}7&\hspace{5pt}0\hspace{5pt}0\hspace{5pt}0\hspace{5pt}0&\hspace{5pt}0\hspace{5pt}4\hspace{5pt} 0\hspace{5pt}0&\hspace{5pt}0\hspace{5pt}4\hspace{5pt}0\hspace{5pt}0\\
0\hspace{5pt}0\hspace{5pt}4\hspace{5pt}0&\hspace{5pt}7\hspace{5pt}1\hspace{5pt}5\hspace{5pt}4&\hspace{5pt}0\hspace{5pt}0\hspace{5pt} 0\hspace{5pt}0&\hspace{5pt}0\hspace{5pt}0\hspace{5pt}4\hspace{5pt}0\\
0\hspace{5pt}0\hspace{5pt}0\hspace{5pt}4&\hspace{5pt}0\hspace{5pt}0\hspace{5pt}0\hspace{5pt}4&\hspace{5pt}4\hspace{5pt}7\hspace{5pt} 1\hspace{5pt}5&\hspace{5pt}0\hspace{5pt}0\hspace{5pt}0\hspace{5pt}0
\end{array}\right).}
\end {eqnarray}

We propose the following theorem for the predicted parity of the Super-Mix function.\begin{table*}[t]\centering
{\centering\caption{{Benchmark for the proposed error detection schemes
for the hash algorithms on ASIC (65nm TSMC)}}
}%
\begin{tabular}{|c|c|c|c|c|c|}
\hline
Algorithm & Block (bits) & {Area {[}GE{]}} & {Frequency {[}MHz{]}} & {Throughput {[}Gbps{]}} & {Efficiency {[}Mbps/GE{]}}\tabularnewline
\hline
\hline
ECHO-256 & \multirow{2}{*}{1,536} & {145,912} & {389} & {6.48} & 44.40\tabularnewline
\cline{1-1} \cline{3-6}
Proposed scheme &  & 187,098 (28\%) & 370 (4.9\%) & 6.18 (4.6\%) & 33.03 (25.6\%)\tabularnewline
\hline
\hline
Fugue-256 & \multirow{2}{*}{32} & {49,040} & {547} & {8.77} & 178.8\tabularnewline
\cline{1-1} \cline{3-6}
Proposed scheme &  & 57,900 (18.1\%) & 519 (5.1\%) & 8.33 (5.1\%) & 141.1 (21.1\%)\tabularnewline
\hline
\end{tabular}
\end{table*}
\begin{theo}
Let ${I_i}\in GF(2^8)$ and ${O_i}\in GF(2^8)$, $0\leq i\leq 15$, be the 16-byte input and output of the Super-Mix function in Fugue, respectively. Then, the predicted parity for this function, i.e., $\hat{P}_{SM}$, is derived as follows (we note that parity is just an example and any other detecting codes can be utilized)
\begin{equation}
\hat{P}_{SM}=\{3\}_h(I_0+I_5+I_{10}+I_{15}),
\end{equation}where the multiplication is performed using the irreducible polynomial $M(x)=x^8+x^4+x^3+x+1$.
\end{theo}
\textit{Proof.}
\textit{
We add the elements of the columns of ${\boldsymbol N}$ to reach the predicted parity $\hat{P}_{SM}$. It is interesting to note that adding the elements in all columns except those in columns $0$, $5$, $10$, and $15$ would result zero. For instance, if one adds the elements in column 1 of ${\boldsymbol N}$ (modulo-2), the result would be $\{4\}_h+\{1\}_h+\{1\}_h+\{7\}_h+\{7\}_h+\{1\}_h+\{5\}_h=0$. For columns $0$, $5$, $10$, and $15$, the addition of elements results in $\{1\}_h+\{4\}_h+\{7\}_h+\{1\}_h=\{4\}_h+\{7\}_h=\{3\}_h$ and this completes the proof. We note that the multiplication with $\{3\}_h=\{2\}_h+\{1\}_h$ is derived by the addition of $I_0+I_5+I_{10}+I_{15}$ with $x(I_0+I_5+I_{10}+I_{15}) \mod M(x)$.}

\section{Simulation Results and ASIC Implementations}
The proposed error detection architectures have
been simulated after injecting faults. The proposed architectures
have the capability of detecting both permanent and transient faults
(this covers both natural and malicious faults). In this paper, we use stuck-at error model. The objective in using this model is to cover the malicious errors injected by the attackers to break the algorithm (by injecting one or more
incorrect bits) and to
detect natural errors caused by bit flips. The stuck-at error forces one bit (for single stuck-at error model) or
multiple bits (for multiple stuck-at error model) to be stuck at logic one
or zero. This makes the result value independent of the
error-free intended value.

In fault attacks, single error injection is the ideal case for gaining the maximum
information. Nevertheless, due to technological constraints, a more realistic error model is to inject multiple errors. Therefore, for covering both natural errors and fault attacks, multiple errors need to be considered. The proposed diagnosis schemes in this paper are independent of
the life-time of errors. Therefore, both permanent and transient
stuck-at errors lead to the same error coverage. We also note that intelligent attackers do not get confined to just multiple stuck-at faults and thus the ability to detect single faults is important.

The fault model used to test the
proposed architectures is created using external feedback linear-feedback
shift registers (LFSRs) to generate pseudo-random fault vectors that
can flip random bits in the output of the gates and at random intervals.
For the architectures presented, we have injected up to 80,000
faults and recorded the number of errors. We have also used the redundant-basis S-boxes in composite field where applicable. Moreover, the false alarm ratios are derived. The error coverage in all the cases is more than 99\% (and for the case of single stuck-at faults, 100\% if we harden the error indication flag comparison units), with relatively low ratio for false alarms, i.e., 0.1\%-0.3\% for the cases. As we inject more faults, the difference
between the error detection results is, comparably, not high, showing the relatively high accuracy of the results.

Through ASIC and for the constructions of the algorithms in 256-bit form, we also present
the performance and implementation metrics of the presented constructions. The benchmarking is performed for the error detection architectures
using TSMC 65nm library and Synopsys Design Compiler (shown in Table
I for area, frequency, throughput, and efficiency {[}throughput over GE{]}). We note that
in Table I, in order to make the area results meaningful when switching
technologies, we have also provided the NAND-gate equivalency (gate
equivalents: GE). This is performed using the area of a NAND gate
in the utilized TSMC 65-nm CMOS library which is 1.41 $\mu m^{2}$. The results presented in Table I show acceptable overhead (degradation) for performance and implementation metrics. We also note that the utilized platform is merely for benchmark and we expect similar results on field-programmable gate arrays (FPGAs) or different ASIC libraries.

\section{Conclusions}
In this paper, we have proposed efficient fault detection schemes by presenting closed formulations for the predicted signatures of different transformations in three hash algorithms. These signatures are derived to achieve low overhead for the specific transformations and can be tailored to include byte/word-wide predicted signatures. Through simulations, we have shown that the proposed fault detection schemes are highly capable of detecting natural hardware failures and are capable of deteriorating the effectiveness of malicious fault attacks. The proposed reliable hardware architectures have been also implemented on ASIC platform using a 65-nm standard technology to benchmark their hardware and timing characteristics. The high efficiency of the proposed schemes makes the proposed reliable architectures suitable for high-performance applications.


\end{document}